\documentstyle[prl,aps,twoside,epsfig]{revtex}
\oddsidemargin=\evensidemargin
\begin{document}
\title{Reexamination of the Constraint on Topcolor-Assisted Technicolor 
Models from $R_b$ }
\author{Chongxing Yue$^{a,b}$,~~~~Yu-Ping Kuang$^{a,c}$,
  Xue-Lei Wang$^{a,c,b}$~~~~Weibin Li$^{b}$}
\vspace{0.2cm}
\address{a.CCAST (World Laboratory) P.O. Box 8730.
 B.J.100080 P.R. of China\\
 b.Department of Physics,Henan Normal University,
 Xinxiang 453002. P.R. of China\\
 c.Department of Physics, Tsinghua University, Beijing 100084,
 P.R. of China \footnote{Mailing address}}
\bigskip\bigskip\bigskip
\date{TUHEP-TH-99109}
\maketitle
\null\vspace{0.5cm}
\begin{abstract}
Recent study on the charged top-pion correction to $R_b$ shows that 
it is negative and large, so that the precision experimental value of $R_b$ 
gives rise to a severe constraint on the topcolor-assisted technicolor models 
such that the top-pion mass should be of the order of 1 TeV. In this paper, we 
restudy this constraint by further taking account of the extended
technicolor gauge boson correction which is positive. With this 
positive contribution to $R_b$, the constraint on the topcolor-assisted 
technicolor models from $R_b$ changes significantly. The top-pion mass is 
allowed to be in the region of a few hundred GeV depending on the models.\\

\end{abstract}

\bigskip\bigskip \bigskip
        
The mechanism of electroweak symmetry breaking remains
an open question in current particle physics despite the success of the 
standard model (SM) tested by the CERN $e^+e^-$ collider LEP and SLAC Large 
Detector (SLD) precision measurement data. In the SM, an elementary Higgs 
field is assumed to be responsible to break
the electroweak symmetry. So far the Higgs bosons has not been found.
Recent investigation shows that the LEP-SLD precision measurement data do not 
really require the existence of a light Higgs boson \cite{BFS}.
Furthermore, theories with elementary scalar fields suffer from the problems of
{\it triviality} and {\it unnaturalness}. To completely avoid these problems
arising from the elementary Higgs field, various kinds of dynamical 
electroweak symmetry breaking mechanisms have been proposed, and among which 
the topcolor-assisted technicolor theory \cite{TC2-1} is an attractive idea.
In the topcolor-assisted technicolor theory, there are two kinds of new heavy 
gauge bosons: (a) the extended technicolor (ETC) gauge bosons including the 
sideways and diagonal gauge bosons, (b) the topcolor gauge bosons including 
the color-octet colorons $C^a_\mu$  and an extra $U(1)$ gauge boson 
$Z^\prime$. The technicolor interactions play the major role in breaking the 
electroweak gauge symmetry and, in addition, give rise to the masses of 
the ordinary leptons and quarks including a very small portion of the 
top-quark mass, namely $\varepsilon m_t$ \cite{m'_t} with a model-dependent 
parameter $\varepsilon\ll 1$. The topcolor interactions also make small 
contributions to the breaking of the electroweak symmetry, and give rise to 
the main part of the top-quark mass $(1-\varepsilon)m_t$ similar to the 
constituent masses of the light quarks in QCD. So that the heaviness of the 
top quark emerges naturally in the topcolor-assisted technicolor theory.
Furthermore, this kind of theory predicts a number of pseudo Goldstone bosons 
(PGBs) including the technipions in the technicolor sector and the top-pion in
the topcolor sector. All the new particles in this theory can
give corrections to the $Z$-pole observables at LEP and SLC, and thus the 
LEP-SLD precision data may give constraints on the parameters in the
topcolor-assisted technicolor theory. These constraints have recently been 
studied in Refs.\cite{BK,LT}. Due to the strong coupling between the top-pion 
and the top and bottom quarks, the top-pion gives rise to a large negative 
correction to the $Z\to b\bar{b}$ branching ratio $R_b$. Together with the 
positive contributions from the colorons and $Z^\prime$, the total topcolor 
correction to $R_b$ is shown to be quite negative which is of the wrong sign 
when comparing the SM value of $R_b$ to the LEP-SLD data. Since the
negative top-pion corrections become smaller when the top-pion is heavier, the 
LEP-SLD data of $R_b$ give rise to certain lower bound on the top-pion mass. 
It is shown in Ref.\cite{BK,LT} that the top-pion mass $m_{\pi_t}$ should not 
be lighter than the order of 1 TeV to make the theory consistent with the 
LEP-SLD data. This implies that the scale of topcolor should be much
higher than what the original model expected \cite{TC2-1}. However, in those 
analyses, the ETC contributions to $R_b$ are not taken into account.
The main ETC corrections to $R_b$ are from the ETC gauge boson contributions.
It has been shown in Ref.\cite{R_b} that the positive diagonal ETC
gauge boson contribution is larger than the negative sideways gauge
boson contribution, and thus the total ETC correction to $R_b$
is positive. It is the purpose of this short paper to investigate how much the 
constraint on the topcolor-assisted technicolor theory from $R_b$ changes when 
this positive ETC correction is included. 

Since the corrections to $R_b$ in the topcolor-assisted technicolor models 
depends on the values of the parameters in the models, we shall consider the 
original topcolor-assisted technicolor model \cite{TC2-1} (it will be referred 
to as Model-I in this paper) and the topcolor-assisted multiscale technicolor 
model \cite{TC2-2} (it will be referred to as Model-II in this paper) as two 
typical examples in the investigation. These two models are different only in 
their ETC parts. For the model-dependent parameter $\varepsilon$, it has been 
shown that the $b\to s\gamma$ rate is sensitive to the value of $\varepsilon$, 
and the CLEO data on the $b\to s\gamma$ rate gives constraint on the value of 
$\varepsilon$, namely $\varepsilon\alt 0.1$ \cite{Balaji}. We shall take three 
values $\varepsilon=0.05,~0.08,$ and 0.1 in our calculation to see its effect.
 
The left-handed and right-handed $Z-b-\bar{b}$ and $Z-t-\bar t$ coupling 
constants $g^b_L,~g^b_R,~g^t_L$ and $g^t_R$ in the SM  are, respectively, 
$g^b_L=-\frac{1}{2}+\frac{1}{3}\sin^2\theta_W,~~g^b_R=\frac{1}{3}\sin^2
\theta_W,~~g^t_L=\frac{1}{2}-\frac{2}{3}\sin^2\theta_W,$ and 
$g^t_R=-\frac{2}{3}\sin^2\theta_W$~\footnote{Here we have ignored the
coupling constant $\frac{e}{\sin\theta_W\cos\theta_W}$ which is
irrelevant to $R_b$.}. Let $\delta g^b_L$ and $\delta g^b_R$ 
denote, respectively, the corrections to $g^b_L$ and $g^b_R$ from the
topcolor-assisted technicolor theory. Then the correction to $R_b$ can be 
expressed as 
\begin{eqnarray}                                              
\displaystyle
\frac{{\delta}R_b}{R_b^{SM}}\equiv \frac{R_b-R^{SM}_b}{R^{SM}_b}
=(1-R^{SM}_b)\frac{2[g_L^b{\delta}
g_{L}^b+g_R^b{\delta}g_{R}^b]}{(g_L^b)^2+(g_R^b)^2}\,,
\label{deltaRb}
\end{eqnarray}
where  $R^{SM}_b=0.2158 \pm 0.0002$ is the SM prediction for $R_b$. We shall 
calculate $\delta g^b_L$ and $\delta g^b_R$ from various sectors in Model-I 
and Model-II.

We first consider the topcolor sector which is the same in Model-I and 
Model-II. The Feynman diagrams for the one-loop charged top-pion corrections to
the $Z-b-\bar b$ vertex and the dependence of $-\delta R_b/R^{SM}_b$ on
$m_{\pi_t}$ have been shown in Figs. 1-2 in Ref.\cite{BK}. Compared with the 
charged top-pion contributions, the neutral top-pion contributions are 
suppressed by a factor of $m_b^2/m_t^2$ and thus can be ignored. In 
Ref.\cite{BK}, the effect of the technicolor contribution to the top-quark 
mass $\varepsilon m_t$ is not taken into account (the result in Ref.\cite{BK} 
corresponds to taking $\varepsilon=0$). Taking account of the $\varepsilon$ 
effect, the total one-loop top-pion correction to $R_b$ in the on-shell
renormalization scheme reads \cite{XWL}
\begin{eqnarray}                                             
&&\delta g_{L}^{b(\pi_t)}=\left(\frac{v_\pi}{v_w}\right)^2
\frac{[(1-\varepsilon)m_t]^2V_{tb}^2}{16\pi^2F_{\pi_t}^2}
\{-g^b_L{\bar{B}}_1(-p_b,m_t,m_{\pi_t})
+g^t_R[2{\bar{C}}_{24}^{\ast}
 +{\bar{B}}_0(-k,m_t,m_t)
 -m_{\pi_t}^2C_0^{\ast}(p_b,-k,m_{\pi_t},m_t,m_t)]\nonumber\\
&&\hspace{1.5cm}+g^t_L m_t^2C_0^{\ast}(p_b,-k,m_{\pi_t},m_t,m_t)
+(1-2\sin^2\theta_W){\bar{C}}_{24}(-p_b,k,m_t,m_{\pi_t},
m_{\pi_t})\}\,,
\label{pitL}
\\
&&{\delta}g_{R}^{b(\pi_t)} = 0\,,                            
\end{eqnarray}
where
$v_\pi/v_w=(167~{\rm GeV})/(174~{\rm GeV})$ reflects the
effect of the mixing between the top-pion and the would-be Goldstone
boson \cite{BK,LT}, $F_{\pi_t}=50$ GeV is the top-pion decay constant, $p_b$ 
and $k$ are, respectively, the momenta of the external $b$ quark and $Z$ 
boson, $B_i$ and $C_{ij}$ are the two-point and three-point scalar integral 
functions. The factor $[(1-\varepsilon)m_t]^2$ in (\ref{pitL}) comes from the 
$\pi_t-t-\bar{b}$ coupling when the technicolor contribution to the top-quark 
mass is considered. This factor causes the $\varepsilon$-dependence of $\delta
g_L^{b(\pi_t)}$. The negative correction to $R_b$ from the top-pion decreases
with $\varepsilon$. In Fig. 1, we plot the top-pion contributed 
$-\delta R_b/R^{SM}_b$ versus $m_{\pi_t}$ with $\varepsilon=0,~0.05,~0.08,
~0.1$. The $\varepsilon=0$ curve is just the result given in Ref.\cite{BK}. 

The contributions to $\delta g^b_L$ and $g^b_R$ from the topcolor gauge bosons 
$C^a_\mu$ and $Z^\prime_\mu$ have been calculated in Ref.\cite{HZ,LT}, which 
are
\begin{eqnarray}                                               
&&\delta g_{L}^{b(C^a)}=g^b_L\frac{\kappa_3 }{6\pi}
C_2(R)\left[\frac{m_Z^2}{M_C^2}\ln\frac{M_C^2}{m_Z^2}\right],\;\;\;\;
\delta g_{R}^{b(C^a)}=g^b_R\frac{\kappa_3 }{6\pi}C_2(R)\left[
\frac{m_Z^2}{M_C^2}\ln\frac{M_C^2}{m_Z^2}\right]\,,
\label{C^a}\\
&&\delta g_{L}^{b(Z^\prime)}=g^b_L\frac{\kappa_1}{6\pi}(Y_L^b)^2
\left[\frac{m_Z^2}{M_{Z'}^2}\ln\frac{M_{Z'}^2}{m_Z^2}\right],\;\;\;\;
\delta g_{R}^{b(Z^\prime)}=g^b_R\frac{\kappa_1}{6\pi}(Y_R^b)^2
\left[\frac{m_Z^2}{M_{Z'}^2}\ln\frac{M_{Z'}^2}{m_Z^2}\right]\label{Z'}\,,
\end{eqnarray}
where $\kappa_3$ and $\kappa_1$ are, respectively, the coloron and the
$Z^\prime$ couplings \cite{TC2-1,LT}, $M_C$ and $M_{Z^\prime}$ are, 
respectively, the masses of $C^a_\mu$ and $Z^\prime$, $C_2(R)=\frac{4}{3}$, 
$Y^b_L=\frac{1}{3}$, and $Y^b_R=-\frac{2}{3}$. We shall take 
$M_B=M_{Z'}=1$ TeV in the calculation. To obtain proper vacuum tilting
(the topcolor interactions only condense the top quark but not the
bottom quark), the couplings $\kappa_3$ and $\kappa_1$ should satisfy
certain constraint. There is a region of $\kappa_3$ and
$\kappa_1$, namely $\kappa_3=2,\;\kappa_1\leq 1$, satisfying the requirement 
of vacuum tilting and the constraints from Z-pole physics and $U(1)$ 
triviality shown in Refs.\cite{BBHK,PS}. We shall take $\kappa_3=2$ and 
$\kappa_1=1$ in the following calculation.

Next, we consider the ETC sector corrections to $R_b$. In the topcolor-assisted
technicolor theory, the technipion-top-bottom coupling is proportional to 
$\frac{\varepsilon m_t}{F_\pi}$, and the technipion corrections to
$\delta g^b_L$ and $\delta g^b_R$ are proportional to 
$(\frac{\varepsilon m_t}{F_\pi})^2$ which is very small, so that the 
technipion correction to $R_b$ is negligible. Therefore, the main 
contribution is from the ETC gauge bosons. This has been calculated in 
Ref.\cite{CSS,Wu,R_b} which reads
\begin{eqnarray}                                                   
{\delta}g_{L}^{b(ETC)}=-\frac{1}{A}\frac{\varepsilon m_{t}}{16\pi F_\pi}
 \sqrt{\frac{N_{TC}}{N_{C}}}[\frac{2N_{C}}{N_{TC}+1}\xi_{t}(
\xi_{t}^{-1}+\xi_{b})-\xi_{t}^{2}]\,,\label{ETC}
\end{eqnarray}
where $N_{TC}$ and $N_{C}$ are, respectively, the number of technicolors
and the numbers of ordinary colors, $\xi_{t}$ and $\xi_{b}$ are
coupling coefficients with $\xi_b=(m_s/m_c)\xi_t^{-1}$ \cite{YKL}, and 
$\xi_t$ is ETC gauge-group dependent. Following  Refs.\cite{CSS,Wu}, we take 
$\xi_t=1/\sqrt{2}$. The factor $\frac{1}{A}$ reflects the walking effect in 
the ETC sector which is taken to be $A=1.7$ in Refs.\cite{YKL,WTC}. The decay 
constant $F_{\pi}$ is different in Model-I and Model-II. In Model-I,
the ETC sector is the one-family ETC model. Considering the mixing
between the top-pion and the would-be Goldstone boson, we have
$N_d(F_\pi/\sqrt{2})^2+F^2_{\Pi_t}=v_w^2$ ($N_d=4$ is
the number of $SU(2)$ doublets in the one-family technifermion sector),
and thus $F_{\pi}=118$ GeV. In Model-II, the ETC sector is the
multiscale technicolor model in which $F_{\pi}=40$ GeV \cite{WTC}. This
difference makes the ETC corrections to $R_b$ very different in Model-I
and Model-II. This positive ETC correction to $R_b$ is larger in
Model-II than in Model-I.

Finally, we add all the corrections together and obtain the total corrections

\begin{eqnarray}                                                
\delta g_{L}^b&=& \delta g_L^{b(\pi_t)}+\delta g_L^{b(C^a)}+\delta
g_L^{b(Z^\prime)}+\delta g_L^{b(ETC)}\,,\label{Ltotal}\\
\delta g_{R}^b&=&\delta g_R^{b(C^a)}+\delta g_R^{b(Z^\prime)}\,,\label{Rtotal}
\end{eqnarray}
in which $\delta g_L^{b(\pi_t)}$, $\delta g_L^{b(C^a)}$, $\delta g_R^{b(C^a)}$,
$\delta g_L^{b(Z^\prime)}$, $\delta g_R^{b(Z^\prime)}$, and 
$\delta g_L^{b(ETC)}$ are given in eqs.(\ref{pitL}),(\ref{C^a}),(\ref{Z'}), 
and (\ref{ETC}), respectively. Plugging (\ref{Ltotal}) and
(\ref{Rtotal}) into (\ref{deltaRb}), we obtain the total correction
$\delta R_b/R^{SM}_b$ and the predicted $R_b=R^{SM}_b+\delta R_b$ in Model-I 
and Model-II. 

Before presenting the numerical results, let us make an examination of
the parameter range $\varepsilon=0.05\--0.1$ which we take in the calculation.
It has been noticed that the ETC sector not only gives rise to a
positive contribution to $R_b$ but also contributes positive correction to
the oblique correction parameter $T$ (or equivalently $\Delta \rho$,
$\Delta \rho=\alpha T$) \cite{PT} due to the violation of the custodial 
symmetry $SU(2)_c$ in the ETC sector \cite{Wu}. In the original ETC model, the 
top quark mass is completely generated by the ETC dynamics, so that the 
violation of $SU(2)_c$ in the ETC sector is very serious and the positive 
contribution to $T$ (or $\Delta \rho$) is so large that it can barely be 
consistent with the experiment \cite{Wu}. Now we examine this problem in the
topcolor-assisted technicolor models. In the topcolor-assisted technicolor 
models, the ETC sector only gives rise to a very small portion of the top 
quark mass, $\varepsilon m_t$, therefore the violation of $SU(2)_c$ in the ETC 
sector is significantly smaller depending on the values of $\varepsilon$. 
It has been shown that the most dangerous positive contrbution to $T$ in the 
ETC sector is from the exchange of the diagonal ETC gauge boson whose 
couplings to the up-type and down-type techniquarks are different, and this 
has been studied in Ref.\cite{Wu}. Since the four-fermion operators
contributing the positive correction to $T$ is also related to the ETC
generation of the top and bottom quark masses, the formula in
Ref.\cite{Wu} can be further expressed in terms of the parameters
$\varepsilon m_t,~\xi_t$ and $\xi_b$ as follows
\begin{eqnarray}                                         
 T^{ETC}=\frac{1}{A}\frac{1}{16\sin^2\theta_W\cos^2\theta_W}
\frac{N_C^2}{N_{TC}(N_{TC}+1)}\frac{\varepsilon m_tF_\pi}{m_Z^2}
\sqrt{\frac{N_{TC}}{N_C}}[\xi_t^{-1}-\xi_b]^2\,.
\end{eqnarray}
For $\varepsilon=0.05,~0.08$ and $0.1$, the values of $T^{ETC}$ are 
0.006, 0.009 and 0.01, respectively. These are to be compared with the 
experimental value $T=0.00\pm 0.15$ \cite{Erler}. We see that, for the 
parameter range $\varepsilon=0.05\--0.1$ which we take in this paper, the ETC 
contributed positive correction to $T$ is small enough to make the theory 
consistent with the experiment \footnote{The corrections to $T$
from the exchange of topcolor gauge boson has been studied in Ref.\cite{CDT}.}.

Now we compare our predicted $R_b$ with the experimental value 
$R^{expt}_b=0.21642\pm 0.00073$ \cite{ex} to get the new constraints on 
the two typical topcolor-assisted technicolor models. The results of
the predicted $R_b$ in Model-I with $\varepsilon=0.05,~0.08,$ and $0.1$ are 
plotted in Fig. 2 together with the experimental value $R^{expt}_b$. The 
horizontal solid line denotes the central value $R^{expt}_b$, and the 
horizontal dotted lines indicate the $1\sigma$ and $2\sigma$ deviations. We 
see from Fig. 1 that the $2\sigma$ constraints on Model-I are
\begin{eqnarray}                                             
&&\varepsilon=0.05:~~~~~~~~~~400~{\rm GeV}\alt m_{\pi_t},\nonumber\\
&&\varepsilon=0.08:~~~~~~~~~~340~{\rm GeV}\alt m_{\pi_t}\alt 900~{\rm GeV},\nonumber\\
&&\varepsilon=0.1:~~~~~~~~~~~~280~{\rm GeV}\alt m_{\pi_t}\alt 770~{\rm GeV}.
\end{eqnarray}
The results of the predicted $R_b$ in Model-II with $\varepsilon=0.05,~0.08$ 
and $0.1$ are plotted in Fig. 3 together with the experimental value 
$R^{expt}_b$ and the $1\sigma$ and $2\sigma$ deviations. Fig. 3 shows that the 
$2\sigma$ constraints on Model-II are
\begin{eqnarray}                                             
&&\varepsilon=0.05:~~~~~~~~~~350~{\rm GeV}\alt m_{\pi_t}\alt 900~{\rm GeV},\nonumber\\
&&\varepsilon=0.08:~~~~~~~~~~250~{\rm GeV}\alt m_{\pi_t}\alt 560~{\rm GeV},\nonumber\\
&&\varepsilon=0.1:~~~~~~~~~~~~220~{\rm GeV}\alt m_{\pi_t}\alt 430~{\rm GeV}.
\end{eqnarray}
We see that the constraints on Model-I and Model-II are different due
to the different values of $F_\pi$ in the two models. Since $F_\pi$
takes a smaller value (causing a larger positive ETC correction to
$R_b$) in Model-II, the allowed top-pion mass is lower in Model-II than in 
Model-I.

From Fig. 2 and Fig. 3, we see that, when the positive ETC gauge boson 
correction to $R_b$ is taken into account, the constraints on the two
typical topcolor-assisted technicolor models are significantly different from 
that shown in Refs.\cite{BK,LT}. As is mentioned in Ref.\cite{BK}, this kind of
constraint should only be regarded as a rough estimate since the
$\pi_t-t-\bar b$ coupling is so strong that higher order corrections
from the top-pion are expected to be important. Anyway, the conclusion of the 
present investigation is that, to be consistent with the
experimental value $R^{expt}_b$, the top-pion mass is roughly in a region of
a few hundred GeV, and thus the scale of topcolor is likely to be around a 
couple of TeV which is not much higher than what is expected in the original 
topcolor-assisted technicolor theory \cite{TC2-1}.

\begin{center}
{\bf Acknowledgment}
\end{center}

This work is supported by the National Natural Science
Foundation of China, the Fundamental Research Foundation of Tsinghua 
University, a special grant from the Ministry of Education
of China, and the Natural Science Foundation of the Henan Scientific 
Committee.

\begin{figure}
\vspace*{-3cm}
\epsfxsize 150mm \centerline{\epsfbox{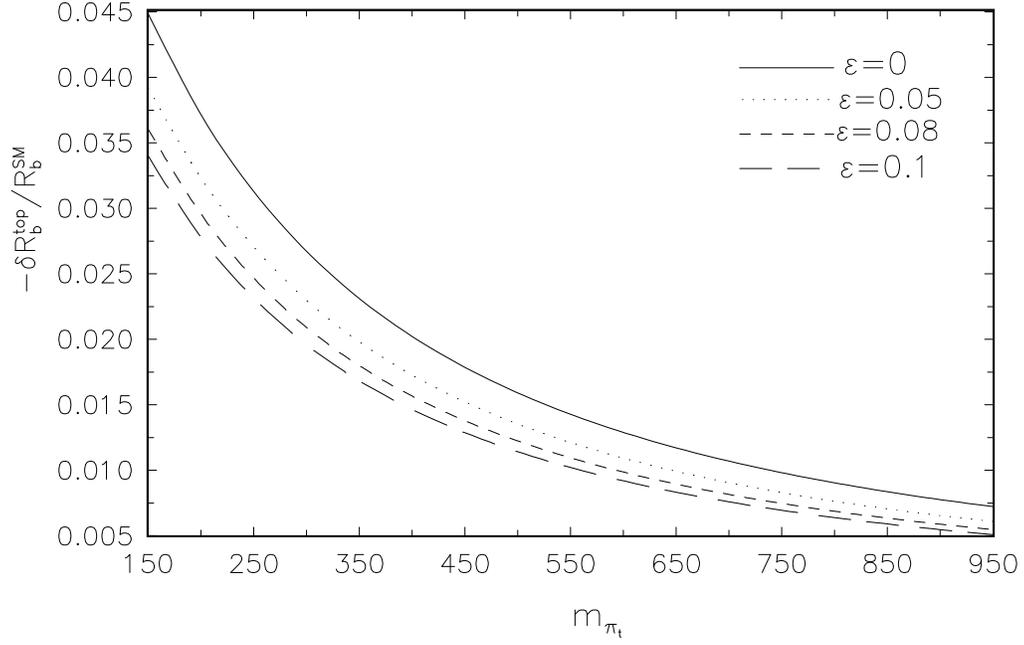}}
\caption{The top-pion contributed $-\delta R_b/R^{SM}_b$ versus the
top-pion mass $m_{\pi_t}$ (in GeV) for $\varepsilon=0,\;0.05,\;
0.08,\;0.1$. }
\end{figure}

\vspace{2cm}
\begin{figure}
\vspace*{-2.6cm}
\epsfxsize 150mm \centerline{\epsfbox{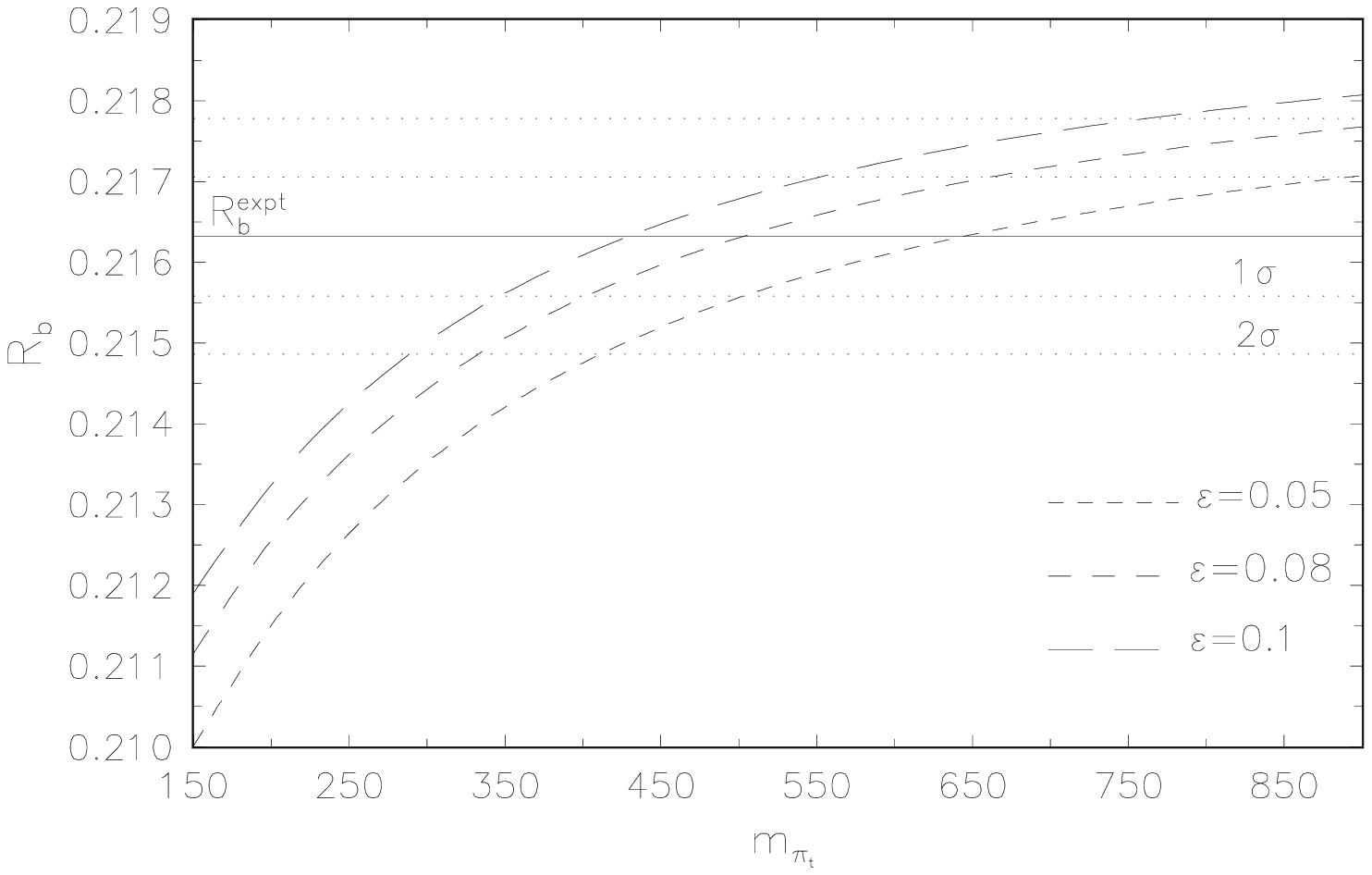}}
\caption{The predicted $R_b$ in Model-I versus the
top-pion mass $m_{\pi_t}$ (in GeV) for $\varepsilon=0.05,\;
0.08,\;0.1$ together with the experimental value $R^{expt}_b$. The horizontal 
solid line denotes the central value of $R^{expt}_b$ and the dotted lines 
show the $1\sigma$ and $2\sigma$ bounds.}
\end{figure}

\begin{figure}
\epsfxsize 150mm \centerline{\epsfbox{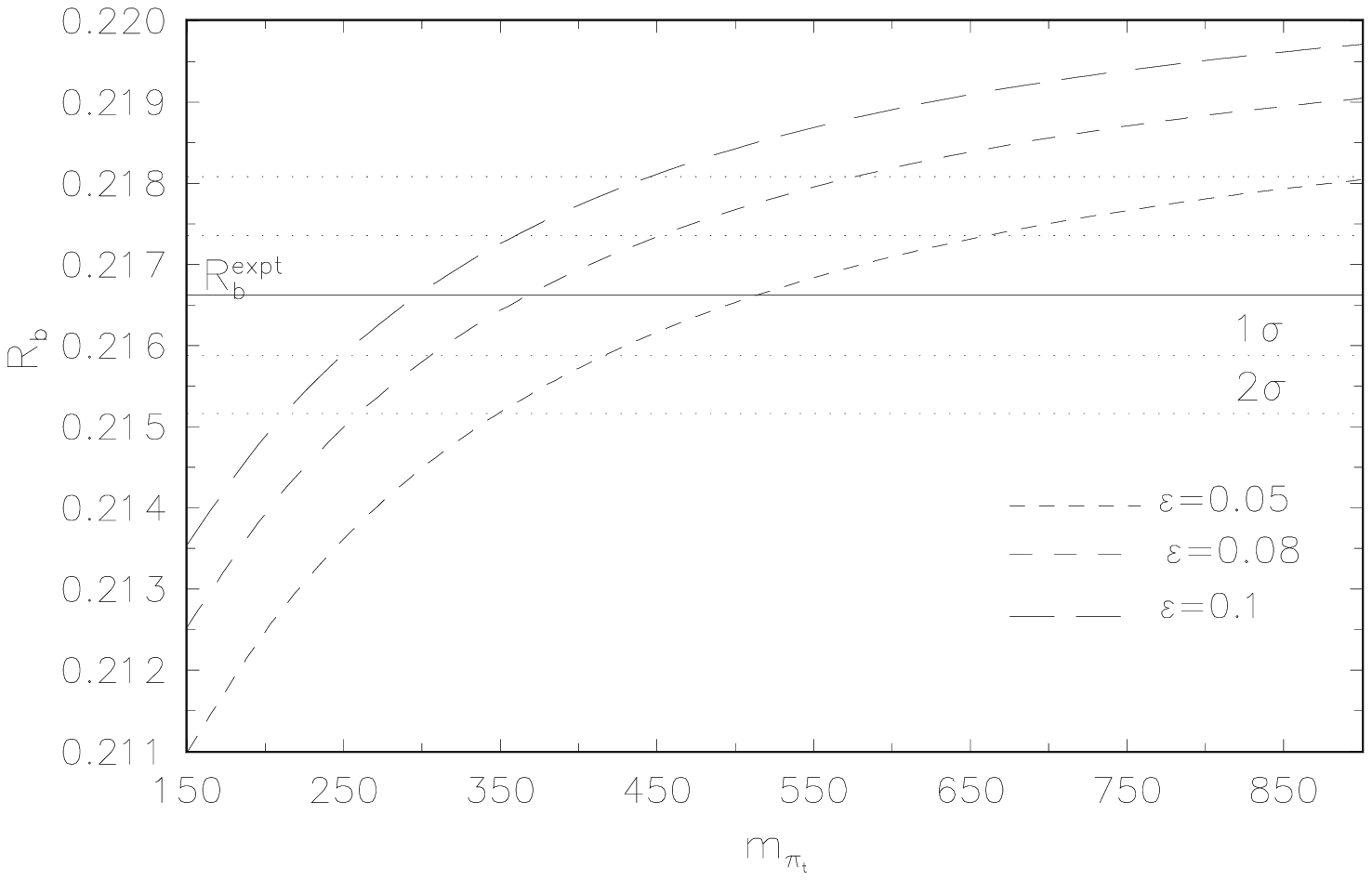}}
\caption{The predicted $R_b$ in Model-II versus the
top-pion mass $m_{\pi_t}$ (in GeV) for $\varepsilon=0.05,\;
0.08,\;0.1$ together with the experimental value $R^{expt}_b$. The horizontal 
solid line denotes the central value of $R^{expt}_b$ and the dotted lines 
show the $1\sigma$ and $2\sigma$ bounds.}
\end{figure}
\end{document}